\documentclass[a4paper,12pt]{article}
\usepackage{amssymb,amsmath,bm}
\usepackage{mathrsfs}
\usepackage{graphicx}
\usepackage{enumerate}

\title{A Note on the Continuity of Expected Utility Functions}
\author{Yuhki Hosoya\thanks{TEL: +81-90-5525-5142, E-mail: ukki(at)gs.econ.keio.ac.jp}\\ Faculty of Economics, Chuo University\thanks{742-1, Higashinakano, Hachioji-shi, Tokyo, 192-0393, Japan.}}
\date{}
\begin{document}
\maketitle

\begin{abstract}
In this paper, we study the continuity of expected utility functions, and derive a necessary and sufficient condition for a weak order on the space of simple probabilities to have a continuous expected utility function. We also verify that almost the same condition is necessary and sufficient for a weak order on the space of probabilities with compact-support to have a continuous expected utility function.

\vspace{12pt}

\vspace{12pt}
\end{abstract}

\section{Introduction}
von Neumann and Morgenstern (1944) presented a necessary and sufficient condition for a weak order on the space of simple probability measures on a set $X$ to be represented in terms of the greater or lesser expected value of a function $u:X\to \mathbb{R}$. This $u$ is called an expected utility function. The expected utility function has a wide range of applications in economics and other social sciences, and is still actively studied.

Neumann--Morgenstern's result does not require any topology on the set $X$. However, the most important application of this result is when $X$ is a half-line in $\mathbb{R}$, which is naturally a topological space. It is natural that the continuity of $u$ would be of interest. Therefore, several researchers sought conditions for $u$ to be continuous. Grandmont (1972) is a representative result in this research area. However, a serious problem arises: if we try to introduce a natural condition for $u$ to be continuous, $u$ becomes always bounded. When $X$ is a half-line, functions that we would want to treat very naturally, such as $\sqrt{x}$ and $\log x$, are all unbounded, which means that the usual result on the continuity of $u$ cannot be treated in the application.

In this paper, we consider identifying a necessary and sufficient condition for $u$ to be continuous. First, following Neumann--Morgenstern, we consider the space of simple probability measures on $X$. If $X$ is Hausdorff, then these are Borel probability measures. Under this setup, we derive a necessary and sufficient condition for an expected utility function to be continuous. (Theorem 1)

Next, we change the object of consideration to the space of compact-support Borel probability measures. If $X$ is a separable metric space, then the space of simple probability measures is dense in the whole space of Borel probability measures under the weak* topology. Therefore, by a natural extension of Theorem 1, we can still derive a necessary and sufficient condition for an expected utility function to be continuous. (Theorem 2)

The structure of this paper is as follows. In Section 2, we define basic terms and derive some basic results. Section 3 is devoted to the derivation of the main results. In Section 4, we discuss the position of this study in comparison with related studies.

\section{Preliminaries}
Consider that a nonempty set $X$ is given. A subset $\succsim$ of $X^2$ is called a {\bf binary relation} on $X$. A binary relation $\succsim$ on $X$ is called a {\bf weak order} on $X$ if and only if, 1) (completeness) either $(x,y)\in \succsim$ or $(y,x)\in \succsim$ for every $x,y\in X$, and 2) (transitivity) if $(x,y)\in \succsim$ and $(y,z)\in \succsim$, then $(x,z)\in \succsim$. If $\succsim$ is a weak order on $X$, then we write $x\succsim y$ instead of $(x,y)\in \succsim$. Moreover, we write $x\succ y$ if and only if $x\succsim y$ and $y\not\succsim x$, and $x\sim y$ if and only if $x\succsim y$ and $y\succsim x$.

Suppose that $\succsim$ is a weak order on $X$. If a function $u:X\to \mathbb{R}$ satisfies the following relationship:
\[x\succsim y\Leftrightarrow u(x)\ge u(y),\]
then $u$ is said to {\bf represent} $\succsim$.

Next, suppose that $X$ is a Hausdorff topological space. Suppose that $P$ is a regular Borel probability measure on $X$.\footnote{We say that a Borel probability measure $P$ is {\bf regular} if and only if, for any Borel set $A$, $P(A)=\sup\{P(K)|K\mbox{ is compact},\ K\subset A\}$.} Then, it is known that there exists a unique least closed set $F^*$ such that $P(F^*)=1$.\footnote{Let $F^*$ be the set of all $x\in X$ such that for any open neighborhood $U$ of $x$, $P(U)>0$. We can easily show that $F^*$ is closed, and $P(X\setminus F^*)=0$ by the regularity. We also mention that there is a counterexample of $P$ with $P(X\setminus F^*)=1$, where $P$ is not regular. Therefore, regularity of $P$ is needed. Note also that, if $X$ is a separable metric space, then such a phenomenon is prohibited, and thus the regularity assumption is not needed.} We call such an $F^*$ the {\bf support} of $P$. If the support is compact, then we call $P$ {\bf compact-support}, and if the support is finite, then we call $P$ {\bf simple}.

Suppose that $X$ is a Hausdorff topological space, $P$ is a regular Borel probability measure, and $u:X\to \mathbb{R}$ is a Borel measurable function. Let $E_P[u]$ denote the expectation of $u$ with respect to $P$. If $P$ is simple and $\{x_1,...,x_m\}$ is the support of $P$, then
\[E_P[u]=\sum_{i=1}^mP(\{x_i\})u(x_i).\]
Let $\mathscr{P}_X$ be the set of all regular Borel probability measures, $\mathscr{P}_c$ be the set of all compact-support elements in $\mathscr{P}_X$,\footnote{Note that, if $X$ is a metric space, then it is known that any compact-support Borel measure is regular.} and $\mathscr{P}_s$ be the set of all simple elements in $\mathscr{P}_X$. Note that, both $\mathscr{P}_c$ and $\mathscr{P}_s$ are convex. Suppose that $Y\subset X$. Let $\mathscr{P}_Y$ denote the set of all $P\in \mathscr{P}_X$ whose support is included in $Y$. Suppose also that $\succsim$ is a weak order on $\mathscr{P}\subset \mathscr{P}_X$. Then, define $\succsim_Y=\succsim\cap (\mathscr{P}_Y)^2$. Note that, $\succsim_Y$ is a weak order on $\mathscr{P}_Y\cap \mathscr{P}$.

Suppose that $\mathscr{P}$ is a convex subset of $\mathscr{P}_X$, and $\succsim$ is a weak order on $\mathscr{P}$. Then, it is said to be {\bf independent} if and only if for every $P,Q\in\mathscr{P}$, $P\succ Q$ implies that
\[(1-t)P+tR\succ (1-t)Q+tR\]
for every $R\in \mathscr{P}$ and every $t\in [0,1[$. Moreover, it is said to be {\bf mixture continuous} if and only if for every $P,Q,R\in \mathscr{P}$ such that $P\succ Q\succ R$, the following two sets
\[\{t\in [0,1]|(1-t)P+tR\succsim Q\},\ \{t\in [0,1]|Q\succsim (1-t)P+tR\}\]
are closed. If there exists a function $u:X\to \mathbb{R}$ such that $U(P)\equiv E_P[u]$ represents $\succsim$, then we call $u$ a {\bf von Neumann--Morgenstern utility function} of $\succsim$, or abbreviately, an {\bf NM utility function}.

We need the following lemmas.

\vspace{12pt}
\noindent
{\bf Lemma 1}. Suppose that $\mathscr{P}$ is a convex subset of $\mathscr{P}_X$ that includes $\mathscr{P}_s$, and $\succsim$ is a weak order on $\mathscr{P}$. If $\succsim$ has an NM utility function, then $\succsim$ is independent. Moreover, such an NM utility function is unique up to a positive affine transformation.

\vspace{12pt}
\noindent
{\bf Proof}. Suppose that there exists an NM utility function $u$ of $\succsim$. Let $P,Q,R\in \mathscr{P}$ and $P\succ Q$. Then, for any $t\in [0,1[$,
\[(1-t)E_P[u]+tE_R[u]>(1-t)E_Q[u]+tE_R[u],\]
which implies that $(1-t)P+tR\succ (1-t)Q+tR$. Therefore, $\succsim$ is independent.

Suppose that $v$ is another NM utility function of $\succsim$. If $u$ is a constant function, then $v$ is also a constant function, and thus $v$ is a positive affine transformation of $u$. Hence, we assume that there exists $x^*,y^*\in X$ such that $u(x^*)>u(y^*)$. Without loss of generality, we assume that $u(x^*)=1,u(y^*)=0$.

Let $\delta_x$ denote the Dirac measure whose support is $\{x\}$. Choose any $x\in X$. Then,
\[u(x)=u(x)u(x^*)+(1-u(x))u(y^*).\]
If $u(x)\in [0,1]$, then $\delta_x\sim u(x)\delta_{x^*}+(1-u(x))\delta_{y^*}$, and thus $v(x)=(v(x^*)-v(y^*))u(x)+v(y^*)$. If $u(x)>1$, then $\delta_{x^*}\sim \frac{1}{u(x)}\delta_x+\frac{u(x)-1}{u(x)}\delta_{y^*}$, and thus $v(x^*)=\frac{v(x)-v(y^*)}{u(x)}+v(y^*)$, which implies that $v(x)=(v(x^*)-v(y^*))u(x)+v(y^*)$. If $u(x)<0$, then $\delta_{y^*}\sim \frac{-u(x)}{1-u(x)}\delta_{x^*}+\frac{1}{1-u(x)}\delta_x$, and thus $v(y^*)=-\frac{u(x)}{1-u(x)}v(x^*)+\frac{1}{1-u(x)}v(x)$, which implies that $v(x)=(v(x^*)-v(y^*))u(x)+v(y^*)$. Therefore, in any case, we have that $v(x)=(v(x^*)-v(y^*))u(x)+v(y^*)$, as desired. This completes the proof. $\blacksquare$

\vspace{12pt}
\noindent
{\bf Lemma 2}. Suppose that $\succsim$ is a weak order on $\mathscr{P}_s$. Then, the following two statements are equivalent.
\begin{enumerate}[1)]
\item $\succsim$ is independent and mixture continuous.

\item There exists an NM utility function $u:X\to \mathbb{R}$ of $\succsim$.
\end{enumerate}

\vspace{12pt}
Because Lemma 2 is well known, we omit the proof.\footnote{This was proved by von Neumann and Morgenstern (1944). For a modern proof, see Theorem 5.15 of Kreps (1988).}

Suppose that $X$ is a Hausdorff topological space, and $Z=C_b(X)$ be the set of all bounded and continuous real-valued functions on $X$. Let $T^*(P)(u)=E_P[u]$. Then, $T^*$ is a mapping from $\mathscr{P}_X$ into $Z'$. Using this mapping, we can prove the following lemma.

\vspace{12pt}
\noindent
{\bf Lemma 3}. Suppose that $X$ is a Hausdorff topological space. Then, there exists a topology on $\mathscr{P}_X$ such that for any net $(P_{\nu})$, this converges to $P$ if and only if for any bounded and continuous function $u:X\to \mathbb{R}$, the net $(E_{P_{\nu}}[u])$ converges to $E_P[u]$.\footnote{If $X$ is a normal Hausdorff space, then $Z'$ is isomorphic to the space of all regular finitely additive Borel measures, and thus this topology is Hausdorff. See, for example, Theorem IV.6.3 of Dunford and Schwartz (1988). However, if $X$ is not normal, then this topology may not be Hausdorff. In particular, if $X$ is not Tychonoff, then there may be two probability measures $P,Q$ such that $E_P[u]=E_Q[u]$ for any bounded and measurable function $u$, and in this case, this topology cannot be Hausdorff.}

\vspace{12pt}
\noindent
{\bf Proof}. As we stated, let $Z=C_b(X)$ be the space of all bounded and continuous real-valued functions on $X$. For each $u\in Z$, define $L_u:Z'\to \mathbb{R}$ as $L_u(T)=T(u)$. Then, $L_u\in Z''$. Let $\sigma_1$ be the set of all $L_u^{-1}(A)$, where $u\in Z$ and $A$ is an open set of $\mathbb{R}$. Let $\sigma_2$ be the set of all intersections of finite families in $\sigma_1$. Let $\tau$ be the set of all unions of families in $\sigma_2$. Then, $\emptyset,Z'\in \sigma_1$, and thus $\emptyset,Z'\in \tau$. If $(U_i)$ is a family of sets in $\tau$, then for each $i$, there exists a family $(V_{ij})$ of sets in $\sigma_2$ such that $U_i=\cup_jV_{ij}$. Thus, $\cup_iU_i=\cup_{i,j}V_{ij}\in \tau$. Finally, if $U_1,U_2\in \tau$, then for $i\in \{1,2\}$, there exists a family $(V_{ij})$ of sets in $\sigma_2$ such that $U_i=\cup_jV_{ij}$. Then, $V_{1j}\cap V_{2j'}\in \sigma_2$ for each $j,j'$. Therefore,
\begin{align*}
U_1\cap U_2=&~\left(\cup_jV_{1j}\right)\cap \left(\cup_{j'}V_{2j'}\right)\\
=&~\cup_{j,j'}(V_{1j}\cap V_{2j'})\in \tau
\end{align*}
which implies that $\tau$ is a topology on $Z'$.

Now, suppose that $(T_{\nu})$ is a net of $Z'$, and $T\in Z'$. Suppose that $(T_{\nu})$ converges to $T$. Choose any $u\in Z$, and consider a net $(T_{\nu}(u))$. This net is rewritten as $(L_u(T_{\nu}))$. Choose any open set $A\subset \mathbb{R}$ such that $T(u)\in A$. Then, $L_u(T)\in A$. Let $B=L_u^{-1}(A)$. Then, $B\in \tau$, and thus there exists $\nu^*$ such that if $\nu\ge \nu^*$, then $T_{\nu}\in B$. This implies that if $\nu\ge \nu^*$, then $T_{\nu}(u)\in A$. Therefore, $(T_{\nu}(u))$ converges to $T(u)$. Conversely, suppose that for each $u\in Z$, $(T_{\nu}(u))$ converges to $T(u)$. Choose any $U\in \tau$ such that $T\in U$. By the construction of $\tau$, there exists $V\in \sigma_2$ such that $T\in V$ and $V\subset U$. Because of the definition of $\sigma_2$, there exist $u_1,...,u_k\in Z$ and open sets $A_1,...,A_k\subset \mathbb{R}$ such that $V=\cap_{i=1}^kL_{u_i}^{-1}(A_i)$. For each $i$, because $(L_{u_i}(T_{\nu}))$ converges to $L_{u_i}(T)$, there exists $\nu_i$ such that if $\nu\ge \nu_i$, $T_{\nu}(u_i)\in A_i$. Because of the definition of the directed set, there exists $\nu^*$ such that $\nu^*\ge \nu_i$ for all $i\in \{1,...,k\}$. If $\nu\ge \nu^*$, then $T_{\nu}\in V\subset U$, which implies that $(T_{\nu})$ converges to $T$.

Let $\tau'$ be the set of all $(T^*)^{-1}(U)$, where $U\in \tau$ and $T^*:\mathscr{P}_X\to Z'$ is defined as above. Then, $\tau'$ is a topology of $\mathscr{P}_X$. By definition, $(P_{\nu})$ converges to $P$ in $\tau'$ if and only if $(T^*(P_{\nu}))$ converges to $T^*(P)$. As we proved above, $(P_{\nu})$ converges to $P$ if and only if for each $u\in Z$, $(E_{P_{\nu}}[u])$ converges to $E_P[u]$. This completes the proof. $\blacksquare$

\vspace{12pt}
We call this topology the {\bf weak* topology}. If $\mathscr{P}\subset \mathscr{P}_X$, then we also define the weak* topology on $\mathscr{P}$ as the relative topology on $\mathscr{P}$ of the original weak* topology. Suppose that $\succsim$ is a weak order on $\mathscr{P}$. Then, we say that $\succsim$ is {\bf sequentially continuous} if there exists a sequence $(X_n)$ of subsets of $X$ such that, 1) $X_n$ is closed, 2) $X_n$ is included in the interior of $X_{n+1}$, 3) $\cup_nX_n=X$, and 4) $\succsim_{X_n}$ is closed in the weak* topology. 

Note that, if $X_n$ is clopen, then $X_n=X_{n+1}$ is not prohibited. In particular, because $X$ is clopen, if $\succsim$ is closed in the weak* topology, then it is automatically sequentially continuous.

We need the following lemma.

\vspace{12pt}
\noindent
{\bf Lemma 4}. Suppose that $\mathscr{P}$ is a convex subset of $\mathscr{P}_c$, and $\succsim$ is a weak order on $\mathscr{P}$. If $\succsim$ is sequentially continuous, then it is mixture continuous.

\vspace{12pt}
\noindent
{\bf Proof}. Choose any $P,Q,R\in \mathscr{P}$ such that $P\succ Q\succ R$. Because $\mathscr{P}\subset \mathscr{P}_c$, there exists a compact set $C\subset X$ such that the supports of $P,Q,R$ are included in $C$. Choose $(X_n)$ as in the definition of sequential continuity, and let $Y_n$ be the interior of $X_n$. Then, $\cup_nY_n=X$, and thus $(Y_n)$ is an open covering of $C$, which implies that there exists $n$ such that $C\subset Y_n$. Then, $P\succ_{X_n}Q\succ_{X_n}R$, and $\succsim_{X_n}$ is closed with respect to the weak* topology. Because the function $t\mapsto (1-t)P+tR$ is continuous with respect to this topology, we have that the sets
\[\{t\in [0,1]|(1-t)P+tR\succsim_{X_n}Q\},\ \{t\in [0,1]|Q\succsim_{X_n}(1-t)P+tR\}\]
are closed, which is equivalent to the closedness of
\[\{t\in [0,1]|(1-t)P+tR\succsim Q\},\ \{t\in [0,1]|Q\succsim (1-t)P+tR\}.\]
Therefore, $\succsim$ is mixture continuous. This completes the proof. $\blacksquare$

\section{Main Result}
Our first main result is as follows.

\vspace{12pt}
\noindent
{\bf Theorem 1}. Suppose that $X$ is a Hausdorff topological space, and $\succsim$ is a weak order on $\mathscr{P}_s$. Then, there exists a continuous NM utility function $u:X\to \mathbb{R}$ of $\succsim$ if and only if $\succsim$ is independent and sequentially continuous. Moreover, such an NM utility function is unique up to a positive affine transformation.

\vspace{12pt}
To show this theorem, we need the following lemma.

\vspace{12pt}
\noindent
{\bf Lemma 5}. Suppose that $X$ is a Hausdorff topological space, and $\succsim$ is a weak order on $\mathscr{P}_s$. Then, there exists a bounded and continuous NM utility function $u:X\to \mathbb{R}$ of $\succsim$ if and only if $\succsim$ is independent and closed in the weak* topology.

\vspace{12pt}
\noindent
{\bf Proof}. Suppose that $u:X\to \mathbb{R}$ is a bounded and continuous NM utility function of $\succsim$. By Lemma 1, $\succsim$ is independent. Define $v(P,Q)=E_P[u]-E_Q[u]$. Because $u$ is bounded and continuous, $v$ is continuous in the weak* topology. Since $\succsim=v^{-1}([0,+\infty[)$, we have that $\succsim$ is closed, as desired.

Conversely, suppose that $\succsim$ is independent and closed. As we have already noted, any closed weak order is sequentially continuous, and thus $\succsim$ is also sequentially continuous. By Lemma 4, $\succsim$ is mixture continuous, and by Lemma 2, there exists an NM utility function $u:X\to \mathbb{R}$ of $\succsim$.

Hereafter, $\delta_x$ denotes the Dirac measure whose support is $\{x\}$. Suppose that $u$ is not upper semi-continuous at $x$. Then, there exists $\varepsilon>0$ such that for any open neighborhood $U$ of $x$, there exists $x_U\in U$ such that $u(x_U)>u(x)+\varepsilon$. Clearly $(x_U)$ is a net that converges to $x$. Therefore, $(\delta_{x_U})$ converges to $\delta_x$ with respect to the weak* topology. Choose any open neighborhood $V^*$ of $x$, and define $P=(1-t)\delta_x+t\delta_{x_{V^*}}$, where $t>0$ is sufficiently small that $u(x)<E_P[u]<u(x)+\varepsilon$. Then, $\delta_{x_U}\succ P$ for all $U$, but $P\succ \delta_x$, which contradicts the closedness of $\succsim$. Therefore, $u$ is upper semi-continuous. By the symmetrical arguments, we can prove that $u$ is lower semi-continuous, and thus $u$ is continuous.

Suppose that $u$ is not bounded from above. Then, there exists $x^*,x_0\in X$ and a sequence $(x_n)$ such that $u(x_0)>u(x^*)$ and $u(x_n)>u(x^*)+2^n(u(x_0)-u(x^*))$. Define $P_n=(1-2^{-n})\delta_{x^*}+2^{-n}\delta_{x_n}$. Then,
\[E_{P_n}[u]=(1-2^{-n})u(x^*)+2^{-n}u(x_n)>u(x_0)=E_{\delta_{x_0}}[u],\]
which implies that $P_n\succ \delta_{x_0}$ for all $n$. Choose any bounded and continuous function $v:X\to \mathbb{R}$. Then,
\[\lim_{n\to \infty}E_{P_n}[v]=\lim_{n\to \infty}[(1-2^{-n})v(x^*)+2^{-n}v(x_n)]=v(x^*)=E_{\delta_{x^*}}[v],\]
and thus, $(P_n)$ converges to $\delta_{x^*}$ with respect to the weak* topology. However, $\delta_{x_0}\succ \delta_{x^*}$, which contradicts the closedness of $\succsim$. Therefore, $u$ is bounded from above. By the symmetrical arguments, we can prove that $u$ is bounded from below, and thus bounded. This completes the proof. $\blacksquare$

\vspace{12pt}
\noindent
{\bf Proof of Theorem 1}. Suppose that $u$ is a continuous NM utility function of $\succsim$. By Lemma 1, $\succsim$ is independent. Define
\[X_n=u^{-1}([-n,n]), Y_n=u^{-1}(]-n-1/2,n+1/2[).\]
Then, $X_n\subset Y_n\subset X_{n+1}$ and $X=\cup_nX_n$. Because $u$ is continuous, $X_n$ is closed and $Y_n$ is open, which implies that $X_n$ is included in the interior of $X_{n+1}$. Let $u_n$ be the restriction of $u$ to $X_n$. Because $u_n$ is a bounded and continuous NM utility function of $\succsim_{X_n}$, by Lemma 5, we have that $\succsim_{X_n}$ is closed.\footnote{Note that, because $X_n$ is closed, $\mathscr{P}_{X_n}$ can naturally be identified with the set of all regular Borel probability measures on $X_n$, and thus our lemmas are applicable.} Hence, $\succsim$ is sequentially continuous.

Conversely, suppose that $\succsim$ is independent and sequentially continuous. Then, there exists a sequence $(X_n)$ of subsets of $X$ such that $X_n$ is closed, $X_n$ is included in the interior of $X_{n+1}$, $\cup_nX_n=X$, and $\succsim_{X_n}$ is closed with respect to the weak* topology.

By Lemma 4, $\succsim$ is mixture continuous, and by Lemma 2, there exists an NM utility function $u$ of $\succsim$. Let $u_n$ be the restriction of $u$ to $X_n$. Then, $u_n$ is an NM utility function of $\succsim_{X_n}$.

Choose any $x\in X$. Then, there exists $n$ such that $x\in X_{n-1}$. Because $\succsim_{X_n}$ is closed and independent, by Lemma 5, there exists a bounded and continuous NM utility function $v_n$ of $\succsim_{X_n}$. By Lemma 1, $u_n$ is a positive affine transformation of $v_n$, and thus $u_n$ is bounded and continuous. Because the interior of $X_n$ contains $X_{n-1}$, it is an open neighborhood of $x$, and thus $u$ is continuous at $x$. Because $x$ is arbitrary, we conclude that $u$ is continuous. This completes the proof. $\blacksquare$

\vspace{12pt}
If $X$ is a separable metric space, then this result can be extended to $\mathscr{P}_c$.

\vspace{12pt}
\noindent
{\bf Theorem 2}. Suppose that $X$ is a separable metric space, and $\succsim$ is a weak order on $\mathscr{P}_c$. Then, there exists a continuous NM utility function $u$ of $\succsim$ if and only if $\succsim$ is independent and sequentially continuous.

\vspace{12pt}
To prove this, we need an additional lemma.

\vspace{12pt}
\noindent
{\bf Lemma 6}. Suppose that $X$ is a separable metric space, and $\succsim$ is a weak order on $\mathscr{P}$, where $\mathscr{P}$ is a convex subset of $\mathscr{P}_X$ that includes $\mathscr{P}_s$. Then, there exists a bounded and continuous NM utility function of $\succsim$ if and only if $\succsim$ is independent and closed in the weak* topology.

\vspace{12pt}
\noindent
{\bf Proof}. Suppose that $u:X\to \mathbb{R}$ is a bounded and continuous NM utility function. Then, by the same arguments as in the proof of Lemma 5, we can show that $\succsim$ is independent and closed.

Conversely, suppose that $\succsim$ is independent and closed. Let $\succsim'=\succsim\cap (\mathscr{P}_s)^2$. Then, $\succsim'$ is an independent and closed weak order on $\mathscr{P}_s$, and by Lemma 5, there exists a bounded and continuous NM utility function $u:X\to \mathbb{R}$ of $\succsim'$.

Note that, $\mathscr{P}_s$ is dense in $\mathscr{P}_X$ with respect to the weak* topology.\footnote{See Theorem II.6.3 of Parthasarathy (1967).} Choose any $P,Q\in \mathscr{P}$. Because $u$ is continuous and bounded, $E_P[u],E_Q[u]$ are well-defined. If $P\succ Q$, then there exist a sequence $((P_n,Q_n))$ on $\mathscr{P}_s^2$ and $R_1,R_2\in \mathscr{P}_s$ such that $P\succ R_1\succ R_2\succ Q$, $P_n\succ R_1$ and $R_2\succ Q_n$ for all $n$, and $(P_n,Q_n)\to (P,Q)$ as $n\to \infty$.\footnote{Choose $R_1$ sufficiently near to $(2/3)P+(1/3)Q$ and $R_2$ sufficiently near to $(1/3)P+(2/3)Q$.} Because $u$ is bounded and continuous, we have that
\[\lim_{n\to \infty}E_{P_n}[u]=E_P[u],\ \lim_{n\to \infty}E_{Q_n}[u]=E_Q[u].\]
Because $P_n\succ R_1\succ R_2\succ Q_n$, $P_n\succ'R_1\succ'R_2\succ'Q_n$, and thus
\[E_{P_n}[u]>E_{R_1}[u]>E_{R_2}[u]>E_{Q_n}[u],\]
which implies that
\[E_P[u]\ge E_{R_1}[u]>E_{R_2}[u]\ge E_Q[u].\]
Next, suppose that $P\succsim Q$. If there exists $R\in \mathscr{P}$ such that $R\succ P$, then for any $t\in ]0,1[$, by independence,
\[tR+(1-t)P\succ tP+(1-t)P=P,\]
which implies that $tR+(1-t)P\succ Q$. Therefore,
\[tE_R[u]+(1-t)E_P[u]>E_Q[u].\]
Letting $t\to 0$, we obtain $E_P[u]\ge E_Q[u]$. By the same argument, if there exists $R\in \mathscr{P}$ such that $Q\succ R$, then we can show that $E_P[u]\ge E_Q[u]$. Suppose that there is no $R\in \mathscr{P}$ such that either $R\succ P$ or $Q\succ R$. If $P\sim Q$, then $P\sim R$ for all $R\in \mathscr{P}$, which implies that $u$ is a constant function. Therefore, $E_P[u]=E_Q[u]$. If $P\succ Q$, then we have already shown that $E_P[u]>E_Q[u]$. Therefore, in either case, we have that $E_P[u]\ge E_Q[u]$. Hence, $u$ is also an NM utility function of $\succsim$. This completes the proof. $\blacksquare$

\vspace{12pt}
\noindent
{\bf Proof of Theorem 2}. Using Lemma 6 instead of Lemma 5, we can show that if there exists a continuous NM utility function of $\succsim$, then $\succsim$ is independent and sequentially continuous by the same proof.\footnote{Note that, any subset of a separable metric space is also a separable metric space.}

Conversely, suppose that $\succsim$ is independent and sequentially continuous. Let $\succsim'=\succsim\cap (\mathscr{P}_s)^2$. Then, $\succsim'$ is an independent and sequentially continuous weak order on $\mathscr{P}_s$, and by Theorem 1, there exists a continuous NM utility function $u:X\to \mathbb{R}$ of $\succsim'$. Choose $(X_n)$ in the definition of sequential continuity, and let $Y_n$ be the interior of $X_n$. Let $u_n$ be the restriction of $u$ to $X_n$. Then, $u_n$ is a continuous NM utility function of $\succsim_{X_n}'$.

Choose any $P,Q\in \mathscr{P}_c$. Then, there exists a compact set $C$ that includes the supports of $P,Q$. Because $(Y_n)$ is an open covering of $C$, there exists $n$ such that $C\subset Y_n$. Because $\succsim_{X_n}$ is independent and closed, by Lemma 6, there exists a bounded and continuous NM utility function $v_n$ of $\succsim_{X_n}$. For any $P',Q'\in \mathscr{P}_{X_n}\cap \mathscr{P}_s$,
\[E_{P'}[v_n]\ge E_{Q'}[v_n]\Leftrightarrow P'\succsim_{X_n}Q'\Leftrightarrow P'\succsim_{X_n}'Q',\]
which implies that $v_n$ is also an NM utility function of $\succsim_{X_n}'$. By Lemma 1, $u_n$ is a positive affine transformation of $v_n$, and thus,
\[E_P[u]\ge E_Q[u]\Leftrightarrow E_P[u_n]\ge E_Q[u_n]\Leftrightarrow E_P[v_n]\ge E_Q[v_n]\Leftrightarrow P\succsim_{X_n}Q\Leftrightarrow P\succsim Q,\]
which implies that $u$ is also a continuous NM utility function of $\succsim$. This completes the proof. $\blacksquare$

\section{Comparison with Related Results}
A foundational result in this context is Lemma 2, which describes a necessary and sufficient condition for the existence of an NM utility function. This result was proved by von Neumann and Morgenstern (1944), and later refined. We first mention this refinement. First, instead of considering $\mathscr{P}_X$ directly, consider a structure called a mixture set. A pair $(X,h)$ is called a {\bf mixture set} if $h:[0,1]\times X^2\to X$ satisfies the following three conditions. First, $h(1,x,y)=x$. Second, $h(t,x,y)=h(1-t,y,x)$. Third, $h(t,h(s,x,y),y)=h(st,x,y)$. If there is no risk of confusion, we use the notation $tx+(1-t)y$ instead of $h(t,x,y)$. Then, the three requirements can be rewritten as 1) $1x+0y=x$, 2) $tx+(1-t)y=(1-t)y+tx$, and 3) $t(sx+(1-s)y)+(1-t)y=stx+(1-st)y$, which seem to be quite natural. As can be easily verified, any convex subset of $\mathscr{P}_X$ is a mixture set. Therefore, Lemma 2 can be refined using the existence theorem for a real-valued function that represents a weak order in the mixture set. This method was established by Herstein and Milnor (1953). It can therefore be stated that Neumann--Morgenstern's result for the existence of an NM utility function was built on the basis of the convex structure, not the topological one.

As an existence theorem for a real-valued function representing a weak order, Neumann--Morgenstern's theorem is the second oldest after Alt (1936). The most prominent theorem in this context is Debreu's (1954). We shall briefly mention Debreu's theorem. Let $X$ be a separable and connected Hausdorff topological space and let $\succsim$ be a weak order on $X$. Then, $\succsim$ is closed as a subset of $X^2$ if and only if there exists a continuous representation $u:X\to \mathbb{R}$ of $\succsim$. By comparison, Neumann--Morgenstern's result does not require any topological structure on $X$, and thus it is independent of Debreu's result. As a corollary, however, even if $X$ is a topological space, nothing can be said about the continuity of the NM utility function from their result alone.

In several studies, researchers have attempted to search for a natural condition for the NM utility function to be continuous. Grandmont (1972) is one of the most representative ones. He proved our Lemma 6 in the case where $\mathscr{P}$ is closed. He used the approach of functional analysis for the proof. Let us explain the main idea of his proof. First, for $Z=C_b(X)$ in the proof of Lemma 3, if $X$ is a separable metric space, then $\mathscr{P}_X$ can naturally be seen as a subset of $Z'$.\footnote{In particular, if $X$ is locally compact, then by Riesz--Markov--Kakutani's representation theorem, $Z'$ is isometric to the space of all Radon measures.} Grandmont considered $\mathscr{P}$ as a convex subset of $Z'$ and introduced the weak* topology. In this case, since $\mathscr{P}$ is a separable metric space with respect to the usual Prohorov metric, it follows from Debreu's theorem that the weak order $\succsim$ on $\mathscr{P}$ is weak* closed if and only if there exists a weak* continuous representation function $U(P)$. He further applied Herstein--Milnor's result and showed that if, additionally, $\succsim$ is independent, then $U(P)$ can be extended to a weak* continuous linear functional defined on the space of Radon measures. It can be proved from a very elementary proposition on the duality that for such $U$, there must exist a function $u\in Z$ such that $U(P)=E_P[u]$, and thus Lemma 6 has now been proved.

However, the fact that $u\in Z$ means that $u$ must be bounded. A typical application of the theory of expected utility is the case $X=]0,+\infty[$ and $u(x)=\frac{x^{1-\theta}-1}{1-\theta}$. ($\theta>0$ is always assumed, and if $\theta=1$, we consider $u(x)=\log x$.) In this case, $u$ is unbounded for any $\theta>0$, and therefore Grandmont's result is not applicable to this model. In other words, the assumption that $\succsim$ is weak* closed is too strong for standard applications.

This paper attempted to solve this problem by introducing a new condition named sequential continuity. This approach has the advantage that if $\mathscr{P}=\mathscr{P}_s$, then a very weak topological assumption on $X$ is needed. In fact, the only assumption we require in Theorem 1 is that $X$ is a Hausdorff topological space. The reason why the Hausdorff property is necessary is that otherwise an element of $\mathscr{P}_s$ may not be a Borel measure. However, this assumption is extremely weak, and thus it cannot even be proved that $\mathscr{P}_s$ is Hausdorff. If $X$ is normal, then the Hausdorff property of $\mathscr{P}_s$ can be easily proved using Urysohn's lemma. In fact, a weaker assumption is sufficient, i.e., if $X$ is Tychonoff, then $\mathscr{P}_s$ is Hausdorff. However, if $X$ is a topological space that is not even Tychonoff, then there is a possibility of the existence of two probabilities $P,Q\in \mathscr{P}_s$ such that $E_P[v]=E_Q[v]$ for any bounded continuous function $v:X\to \mathbb{R}$. If so, any open set containing $P$ also contains $Q$, and thus $\mathscr{P}_s$ is not Hausdorff. We have carefully checked the seemingly well-known property of weak* topology in Lemma 3 in order to carefully verify that our proof is valid even in such an odd topological space.

Unfortunately, our result can only be discussed up to the case $\mathscr{P}\subset \mathscr{P}_c$. If $P$ were not compact-support, then $E_P[u]$ may not be definable for unbounded $u$. Since our aim is to derive a necessary and sufficient condition for the NM utility function to be continuous, we have to accept this restriction as unavoidable. If some growth condition can be introduced in addition to continuity, then there are several known representation theorems for a continuous NM utility function on spaces including some $P$ that is not compact-support. See, for example, Dillenberger and Krishna (2014).\footnote{Dillenberger--Krishna treat the sets $W_g(X)=\{u:X\to \mathbb{R}|$u$\mbox{ is continuous and }\sup u/g<+\infty\}$ and $\mathscr{P}_g(X)=\{P\in \mathscr{P}_X|E_P[g]<\infty\}$, where $g:X\to [1,\infty[$ is a continuous function. Suppose that $X$ is a separable metric space, $\succsim$ is a weak order on some $\succsim\subset \mathscr{P}\times \mathscr{P}$, and $u:X\to \mathbb{R}$ is a continuous NM utility function of $\succsim$. In this case, using Urysohn's lemma appropriately, we can construct a continuous function $g$ such that $u\in W_g(X)$. Therefore, if $\mathscr{P}\subset \mathscr{P}_g(X)$, then by their Theorem 3.1, $\succsim$ is closed with respect to the weak* topology defined by using $W_g(X)$ instead of $C_b(X)$. However, this topology depends on $g$, and $g$ depends on $u$, which implies that the topology that makes $\succsim$ closed itself depends on $\succsim$. This seems a little odd. Hence, it is difficult to use their result to appropriately axiomatize the continuity of the NM utility function without any additional assumption.}

\section*{References}

\begin{description}
\item{[1]} Alt, F. (1936) ``\"Uber die Messbarkeit des Nutzens.'' Zeitschrift f\"ur National\"okonomie 7, pp.161-169. Translated by Schach, S. (1971) ``On the Measurability of Utility.'' In: Chipman, J. S., Hurwicz, L., Richter, M. K., Sonnenschein, H. F. (Eds.) \textit{Preferences, Utility and Demand}, Harcourt Brace Jovanovich, New York. pp.424-431.

\item{[2]} Debreu, G. (1954) ``Representation of a Preference Ordering by a Numerical Function.'' In: Thrall, R. M., Coombs, C. H., Davis, R. L. (Eds.) \textit{Decision Processes.} Wiley, New York, pp.159-165.

\item{[3]} Dillenberger, D. and Krishna, R. V. (2014) ``Expected Utility without Bounds --- A Simple Proof.'' Journal of Mathematical Economics 52, pp.143-147.

\item{[4]} Dunford, N. and Schwartz, J. T. (1988) \textit{Linear Operators Part 1, Genaral Theory}. Wiley, New York.

\item{[5]} Grandmont, J-M. (1972) ``Continuity Properties of a von Neumann-Morgenstern Utility.'' Journal of Economic Theory 4, pp.45-57.

\item{[6]} Herstein, I. N. and Milnor, J. (1953) ``An Axiomatic Approach to Measurable Utility.'' Econometrica 21, pp.291-297.

\item{[7]} Kreps, D. (1988) \textit{Notes on the Theory of Choice}. Westview Press, Boulder.

\item{[8]} von Neumann, J. and Morgenstern, O. (1944) \textit{Theory of Games and Economic Behavior}. Princeton University Press, Princeton.

\item{[9]} Parthasarathy, K. R. (1967) Probability and Mathematical Statistics. Academic Press, New York.

\end{description}

\end{document}